\documentclass[conference]{IEEEtran}
\IEEEoverridecommandlockouts
\usepackage{cite}
\usepackage{amsmath,amssymb,amsfonts}
\usepackage{algorithmic}
\usepackage{graphicx}
\usepackage{textcomp}
\usepackage{xcolor}
\usepackage[caption=false,font=footnotesize]{subfig}
\usepackage{acronym}
\usepackage{booktabs}
\usepackage{multirow}
\usepackage{siunitx}
\usepackage{url}
\def\BibTeX{{\rm B\kern-.05em{\sc i\kern-.025em b}\kern-.08em
    T\kern-.1667em\lower.7ex\hbox{E}\kern-.125emX}}

\acrodef{VLM}[VLM]{Vision-Language Model}
\acrodef{VQA}[VQA]{Visual Question Answering}
\acrodef{ICM}[ICM]{Image Coding for Machines}

\usepackage{fancyhdr}
\fancypagestyle{arxivheader}{
    \fancyhf{}
    
    \chead{\small \textcolor{gray}{\textit{To appear in the 2026 IEEE International Conference on Multimedia and Expo (ICME 2026)}}}
}

\begin{document}

\title{Prompt-Guided Prefiltering for VLM Image Compression}

\author{
    \IEEEauthorblockN{Bardia Azizian and Ivan V. Baji\'{c}}
    \IEEEauthorblockA{\textit{School of Engineering Science, Simon Fraser University}, Burnaby, BC, Canada \\
    bardia\_azizian@sfu.ca, ibajic@ensc.sfu.ca}

\thanks{This work was supported in part by a gift from Intel Labs, NSERC grant RGPIN-2021-02485, and Digital Research Alliance of Canada.}
}


\maketitle
\thispagestyle{arxivheader}

\begin{abstract}
The rapid progress of large \acp{VLM} has enabled a wide range of applications, such as image understanding and \ac{VQA}. Query images are often uploaded to the cloud, where VLMs are typically hosted, hence efficient image compression becomes crucial. However, traditional human-centric codecs are suboptimal in this setting because they preserve many task-irrelevant details. Existing \ac{ICM} methods also fall short, as they assume a fixed set of downstream tasks and cannot adapt to prompt-driven VLMs with an open-ended variety of objectives. We propose a lightweight, plug-and-play, prompt-guided prefiltering module to identify image regions most relevant to the text prompt, and consequently to the downstream task. The module preserves important details while smoothing out less relevant areas to improve compression efficiency. It is codec-agnostic and can be applied before conventional and learned encoders.
Experiments on several \ac{VQA} benchmarks show that our approach achieves a 25–50\% average bitrate reduction while maintaining the same task accuracy. Our source code is available at \url{https://github.com/bardia-az/pgp-vlm-compression}.
\end{abstract}


\section{Introduction}
\label{sec:intro}
With the rapid advancement of large-scale multimodal foundation models—such as GPT, Gemini, LLaVA, and InternVL series~\cite{gpt5,gemini25,llava_ov,internvl3}—these systems are increasingly being adopted across a wide range of real-world applications. However, due to their massive computational and memory requirements, frontier models cannot be deployed on edge devices and must instead be hosted on the cloud. This creates a strong demand for efficient data transmission from the edge to the cloud, especially for visual data. While textual inputs are typically lightweight, images and videos are much more data-intensive, making visual data compression a critical bottleneck in multimodal analysis systems.

Various Image Coding for Machines (ICM) studies~\cite{harell2025_RDcfm, azizian2024_ppColab} have shown that machine vision tasks typically require far less information from images than human viewing. Human-centric codecs, such as JPEG~\cite{wallace1991jpeg}, HEVC~\cite{HEVC_overview}, and VVC~\cite{VVC_overview}, are optimized for the Human Visual System (HVS) and therefore tend to preserve a lot of information 
irrelevant to a specific task.
\ac{ICM} approaches achieve higher compression efficiency 
by retaining only task-relevant information while discarding irrelevant visual details. However, this design inherently assumes that the downstream task is known in advance, since the encoder must be optimized for that specific task. This assumption breaks down in the context of open-ended, prompt-driven Vision–Language Models (\acp{VLM}), where the downstream task is defined dynamically by the natural-language prompt. Designing a single compression model that is optimized for all possible tasks becomes infeasible unless the encoder also receives prompt information that reveals the nature of the task.

This is precisely the challenge we address in this work. To leverage the prompt and infer task-relevant visual regions, one would ideally deploy a full \ac{VLM} at the encoder side. However, this is impractical for resource-constrained edge devices such as smartphones and drones. Furthermore, since such devices, especially battery-operated ones, typically employ optimized hardware-based image encoders, modifying the codec's internal structures is often infeasible. To address these limitations while adhering to edge device constraints, the main contributions of this work are as follows:

\begin{itemize} 
    \item We introduce a \textbf{prompt-guided prefiltering module} that utilizes TinyCLIP~\cite{tinyclip} to dynamically identify and preserve task-relevant image regions. To the best of our knowledge, this is the first compression framework for \acp{VLM} that 
    adapts to the specific language prompt. 
    \item The proposed method is \textbf{lightweight and codec-agnostic}, designed for seamless edge deployment. It operates as a plug-and-play preprocessor compatible with conventional and learned codecs, requiring no modifications to the encoder or cloud-based decoder. 
    \item Extensive evaluations on two \ac{VLM} backbones and three \ac{VQA} benchmarks demonstrate average bitrate reductions of approximately 28\%, 42\%, 47\%, and 44\% for JPEG, HEVC, VVC, and Cheng's learned image codec~\cite{Cheng_2020_CVPR}, respectively, at the same task accuracy level.
\end{itemize}

\section{Related Work}
\label{sec:rel-work}
The visual input to \acp{VLM} is substantially larger than the language input. This has motivated a line of work aimed at reducing the number of vision tokens to improve the computational and memory efficiency of large \ac{VLM} inference~\cite{rao2021dynamicvit, bolya2023tome, Ye_2025_CVPR, Yang_2025_CVPR}. These approaches are typically referred to as visual token pruning or reduction, some of which leverage guidance from the textual information~\cite{zhu_2024_focusllava}. While our proposed idea could, in principle, be combined with token-pruning methods, our primary goal is to compress the input image itself to enable efficient data transmission in edge-cloud inference pipelines.

On the other hand, studies on compression models tailored specifically for \acp{VLM} are still very limited. The few existing approaches largely aim to build general-purpose models capable of supporting any task defined by the prompt.
These works often prioritize preserving most of the visual information--sometimes nearly at pixel-level fidelity--to remain robust across all potential downstream tasks.

For example, Li \textit{et al.}~\cite{Li_2025_compress_lvlm} propose a variable bit-rate end-to-end learned image compression framework for large \acp{VLM}. Their method introduces a generic, semantics-driven pre-editor that adjusts the input image by emphasizing important regions based on visual tokens, without considering the downstream task or the language prompt. The pre-editor and codec are jointly trained using a loss function that includes rate, pixel-level distortion, and a token-based semantic loss. 
Unlike this jointly-trained pre-editor, our prefiltering module is able to work with any existing codec without modification.

Liu \textit{et al.}~\cite{Liu_2024_tell_codec} propose deploying a resource-heavy Grounded-SAM~\cite{ren2024groundedsam} together with a full large \ac{VLM} on the encoder side to extract semantically structured regions of the image and compress for machine vision tasks. 
However, if an edge device can run such heavy computations locally, compressing the image for cloud inference becomes redundant.

The authors of \cite{kao2025bridging} propose a lightweight transform-neck---a strategy common in \ac{ICM} literature~\cite{Choi_2025_scalable}---to adapt the compressed image latents of a learned codec for \ac{VLM} tasks. This transform-neck bypasses the need to reconstruct the image through the synthesis transform; instead, it directly reconstructs the features at a specific layer of the \ac{VLM}'s vision encoder using a distillation loss. 

A conceptually similar network is introduced in \cite{liu2025_mllms}, referred to as an ``Adapter,'' which extracts semantic features directly from the latent codes. Unlike \cite{kao2025bridging}, this Adapter is added on top of the image decoder and injects high-level guidance into the feature domain, complementing the low-level information obtained through full reconstruction. This design is motivated by the analysis showing that \acp{VLM} rely on visual cues across multiple feature levels, which are unevenly affected by compression artifacts. To address this, the authors introduce a multi-level training objective to improve robustness across the feature hierarchy. Additionally, a shallow CLIP-based~\cite{clip} vision encoder guides generic bit allocation---yet again, without considering the language prompt.

Reference~\cite{Shen_2025_sem_compress} also uses a pretrained CLIP~\cite{clip}, but it encodes only the semantic visual embeddings via a learned codebook optimized to preserve cosine similarity between the reconstructed and original representations, without incorporating any pixel-level reconstruction. While this approach appears effective for high-level classification tasks, its efficacy for complex \ac{VQA} tasks remains unverified, as it has not been evaluated on such benchmarks.

None of the above-mentioned methods considers the fact that the ultimate task in \acp{VLM} is determined by the language prompt, and that the relevant visual information may vary accordingly. Prior studies largely focus on preserving semantically salient regions of the image in a prompt-agnostic manner. As a result, they may fail in scenarios where the task depends on subtle or inconspicuous regions--for example, questions about background details or secondary objects. In contrast, our method explicitly adapts the prefiltering process to the prompt, preserving the fidelity of the prompt-relevant regions while allowing more aggressive compression in irrelevant areas.

\section{Proposed Method}
\label{sec:method}
The primary goal of our proposed method is to leverage information from the prompt to identify the image regions most relevant to the target task and maintain their fidelity, while more aggressively compressing the less relevant areas. To achieve this, we employ TinyCLIP~\cite{tinyclip}, a distilled and lightweight version of CLIP~\cite{clip} that offers comparable performance and can be feasibly deployed on edge devices.\footnote{Apple's and Google's foundation models targeted for the latest generation smartphones range from 3 to 4 billion parameters. Meanwhile, we use TinyCLIP models between 23-120 million parameters in our experiments, well within the processing capability of today's smartphones.} 
See Appendix A for detailed complexity and latency analysis.

\begin{figure*}[htbp]
\centerline{\includegraphics[width=\textwidth]{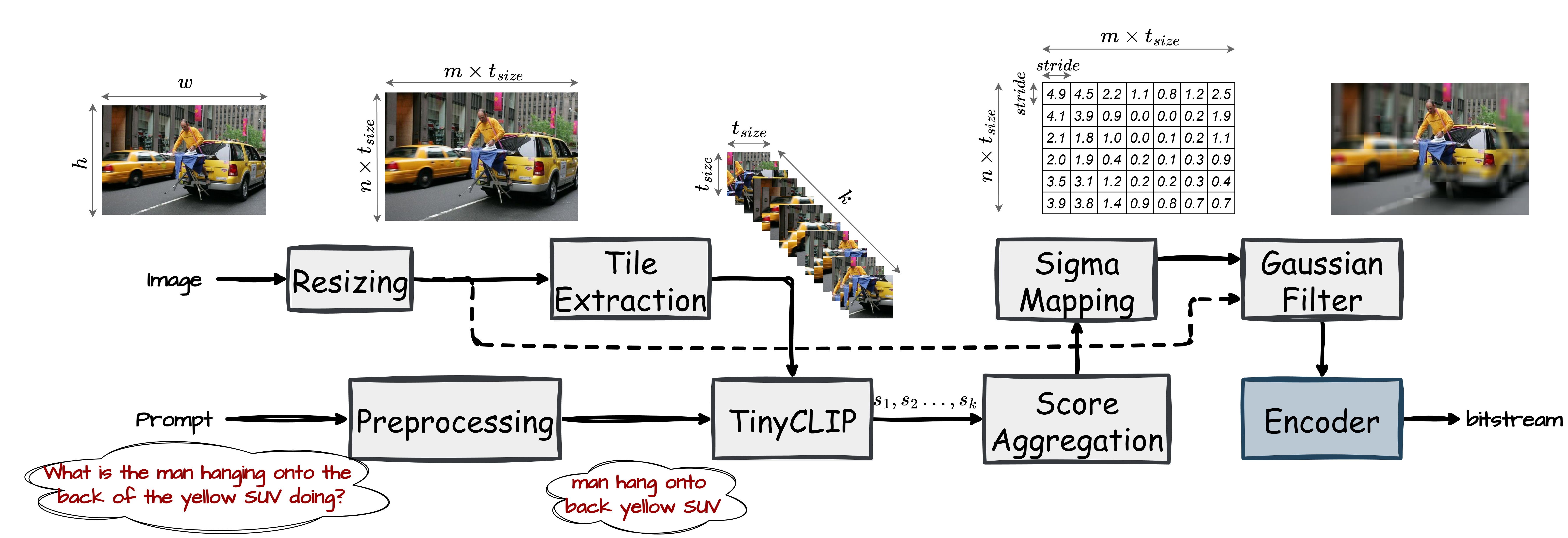}}
\caption{The overall block diagram of the proposed method}
\label{fig:diagram}
\end{figure*}

The overall block diagram of the proposed method is shown in Fig.~\ref{fig:diagram}. After resizing, the input image is divided into several tiles, and the TinyCLIP model assigns a relevance score to each tile based on the preprocessed text it receives. These scores are then aggregated to form a score map. 
Next, the scores are mapped to $\sigma$ values, which parameterize a Gaussian filter. 
Consequently, the filtered image becomes smoother in low-relevance regions to improve compression efficiency, while task-critical areas with higher scores retain finer details.
Finally, the filtered image is compressed using an image encoder of choice. Therefore, the prefiltering module (shown as the light-gray blocks in Fig.~\ref{fig:diagram}) is a plug-and-play component that can precede any image encoder, without the need to change the encoder or decoder. A more detailed description of each component is provided below.

\subsection{Resizing and Tile Extraction}
\label{subsec:tile}
The TinyCLIP model supports a fixed input size of $224 \times 224$. Consequently, the tile size, denoted by $t_{size}$, is also set to $224$ in our scheme. Initially, the input image is resized to the nearest larger multiple of $t_{size}$ in both height and width. The resized image is then divided into tiles using a sliding window, which can be either overlapping or non-overlapping, with a stride applied in both dimensions. The stride value is chosen from $\{t_{size}, t_{size}/2, t_{size}/4\}$ and is automatically selected by checking which of these three possible strides results in a total number of tiles that is closest to the target hyperparameter $\mathtt{tile\_num}$. A larger $\mathtt{tile\_num}$ results in a smaller stride, increasing the overlap between tiles and producing a more granular and accurate score map. However, processing more tiles per image with TinyCLIP increases computational cost, memory usage, and potential latency on edge devices. Therefore, $\mathtt{tile\_num}$ should be carefully selected according to the target application and the capabilities of the edge device.

\subsection{Prompt Preprocessing}
\label{subsec:preproc}
The text input to TinyCLIP does not need to be identical to the prompt used for the main \ac{VLM}. This is because TinyCLIP is not responsible for the downstream task, but rather serves solely to identify image regions most semantically relevant to the input text. Given TinyCLIP's limited context window, the prompt is preprocessed to extract its essential meaning. To this end, non-essential instructional phrases (e.g., ``Answer the question using a single word or phrase'') are first removed. Subsequently, we employ the Natural Language Toolkit (NLTK) library~\cite{bird2009nltk} to discard common stop words (e.g., ``the,'' ``to,'' ``of'') and lemmatize the remaining words to their dictionary forms (e.g., ``running'' $\rightarrow$ ``run'').

Following this step, if the token count still exceeds the context window, the prompt is further refined via part-of-speech analysis. Priority is assigned based on word type, with nouns receiving the highest precedence, followed by adjectives, verbs, and finally other parts of speech. Low-priority tokens are iteratively pruned until the sequence fits within TinyCLIP’s context window.

\subsection{TinyCLIP}
\label{subsec:clip}
As mentioned earlier, TinyCLIP is relatively small compared to its parent model, CLIP. Different TinyCLIP configurations range from $23$M to $120$M parameters. The choice of the model should be guided by the target application and the resources available on the edge device.

TinyCLIP encodes the text and the $i$-th tile ($i = 1, \ldots, k$) into feature embeddings $\mathbf{e}_{\text{text}}$ and $\mathbf{e}_{\text{tile}_i}$, represented as column vectors. Each embedding is then normalized by its $\ell_2$ norm, resulting in the unit vectors:
\begin{equation}
\hat{\mathbf{e}}_{\text{text}} = \frac{\mathbf{e}_{\text{text}}}{\|\mathbf{e}_{\text{text}}\|_2}, \quad
\hat{\mathbf{e}}_{\text{tile}_i} = \frac{\mathbf{e}_{\text{tile}_i}}{\|\mathbf{e}_{\text{tile}_i}\|_2}.
\end{equation}
By applying the dot product between the text and tile embeddings, a similarity score is obtained for each tile:
\begin{equation}
    \mathbf{s} = \text{softmax} \Big( \mathtt{logit\_scale} \cdot (\hat{\mathbf{e}}_{\text{text}}^\top \hat{\mathbf{E}}_{\text{image}}) \Big) \cdot k,
    \label{eq:score-cal}
\end{equation}
where $\hat{\mathbf{E}}_{\text{image}} = [\hat{\mathbf{e}}_{\text{tile}_1}, \hat{\mathbf{e}}_{\text{tile}_2}, \ldots, \hat{\mathbf{e}}_{\text{tile}_k}]$ is the image embedding matrix formed by concatenating the normalized tile embeddings as column vectors. The hyperparameter $\mathtt{logit\_scale}$ controls the sharpness of the resulting score distribution produced by the softmax function. The outputs of the softmax are also multiplied by the number of tiles ($k$) to ensure the sum of the scores is $k$ (and thus the mean score is 1), making the score values independent of the total number of tiles. In other words, if all tiles are equally important for the target task and receive the same score, the resulting scores would be $1$ for all tiles. This normalization is useful for the subsequent Sigma Mapping step (Section \ref{subsec:filter}). Finally, the resulting vector $\mathbf{s}$ contains the score for each tile (i.e. $\mathbf{s} = [s_1, s_2, \ldots, s_k]$).

\subsection{Score Aggregation}
\label{subsec:score}
In the Score Aggregation block, the tile-level scores are merged such that pixels belonging to multiple overlapping tiles receive the average of those tiles’ scores. Since the tile size ($t_{size}$) is always an integer multiple of the \textit{stride}, the sliding windows align perfectly at their boundaries, forming a regular grid of size $\textit{stride} \times \textit{stride}$ across the image. Consequently, all pixels within each $\textit{stride} \times \textit{stride}$ region share the same aggregated score value.

\subsection{Sigma Mapping and Gaussian Filter}
\label{subsec:filter}
A Gaussian filter is a linear smoothing operation that convolves the image with a Gaussian kernel (fixed to a size of $11 \times 11$ in our implementation). We employ this low-pass filter because it is the only circularly-symmetric filter that allows for fast separable (row-column) implementation, and its bandwidth is easy to control with a single parameter ($\sigma$). 
Furthermore, unlike other smoothing kernels, the Gaussian filter guarantees that no spurious structures (artifacts) are introduced into the image during the smoothing process~\cite{babaud1986uniqueness}. This ensures that the visual semantics remain authentic for the VLM, preventing the generation of misleading features.

The standard deviation ($\sigma$) of the kernel determines the filter’s bandwidth and, consequently, the degree of smoothing. Larger $\sigma$ values result in stronger smoothing. Therefore, $\sigma$ is designed to vary inversely with the relevance scores obtained from the TinyCLIP model. To model this inverse relationship, we employ an exponential mapping from the score values to the corresponding $\sigma$ values:
\begin{equation}
\sigma = \sigma_{\mathbf{1}} \left( \frac{\sigma_{\max}}{\sigma_{\mathbf{1}}} \right)^{(1 - \text{score})}
\label{eq:exp_mapping}
\end{equation}
where both $\sigma_{\max}$ and $\sigma_{\mathbf{1}}$ are treated as hyperparameters. $\sigma_{\max}$ denotes the maximum $\sigma$ value (occurring at $\text{score} = 0$), while $\sigma_{\mathbf{1}}$ represents the $\sigma$ value at $\text{score} = 1$, which is the theoretical mean score resulting from the normalization factor $k$ in \eqref{eq:score-cal}. For scores greater than $1$, $\sigma$ decreases rapidly toward zero. We empirically evaluated alternative mapping functions, including linear and reciprocal variants, and found that the exponential mapping~(\ref{eq:exp_mapping}) yielded superior performance.

Finally, a Gaussian filter parameterized by the corresponding $\sigma$ is applied locally to each $\textit{stride} \times \textit{stride}$ block, using reflection padding at the boundaries. After prefiltering, the image is encoded using the codec of choice, then decoded, and then fed to the VLM together with the original prompt.

\section{Experiments}
\label{sec:exp}
In our testing setup, the model in Fig.~\ref{fig:diagram} generates a compressed bitstream from the input image. The decoder then reconstructs the image and forwards it to the main \ac{VLM} for analysis together with the original (non-preprocessed) prompt. 

\begin{figure}
    \centering
    \includegraphics[width=\columnwidth]{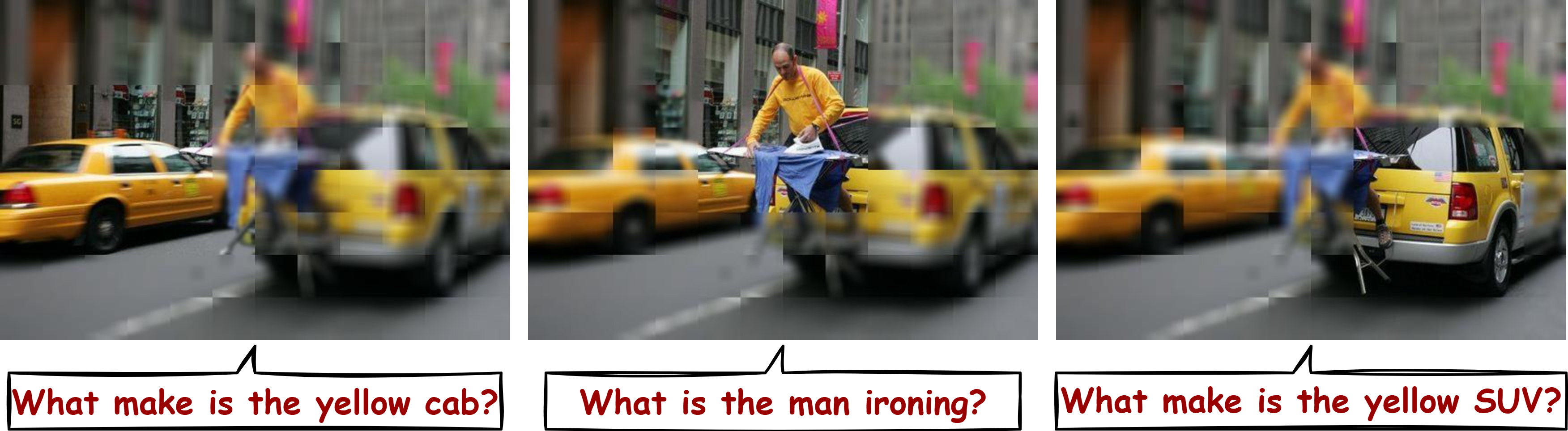}
    \caption{Visualization of the prefiltered versions of a single image generated by our method under three different prompts.}
    \label{fig:vis}
\end{figure}

Fig.~\ref{fig:vis} shows filtered outputs for three different prompts. As illustrated, the same image is processed differently depending on the prompt-specific region of interest. With the help of TinyCLIP, the module identifies these important regions, preserves their details, and smoothens task-irrelevant areas, resulting in improved compression efficiency.

\subsection{Compression Efficiency}
\label{subsec:compress_eff}
In our experiments, we used a TinyCLIP model instance\footnote{\texttt{TinyCLIP-auto-ViT-22M-32-Text-10M}} with a total of 57M parameters. We set $\mathtt{tile\_num} = 24$, $\mathtt{logit\_scale} = 20$, $\sigma_{\mathbf{1}} = 0.2$, and $\sigma_{\max} = 3.0$.

To evaluate the compression efficiency of our method, we conducted experiments on three \ac{VQA} benchmarks: MME~\cite{fu2023mme}, MMBench~\cite{liu2024mmbench}, and SEEDBench~\cite{Li_2024_CVPR} using VLMEvalKit~\cite{duan2024vlmevalkit}. To demonstrate the generalizability of our approach, we compared systems relying solely on JPEG, HEVC, VVC, or the learned codec by Cheng \textit{et al.}~\cite{Cheng_2020_CVPR} against the same pipelines augmented with our prefiltering module. JPEG and HEVC compression were implemented using the Pillow library and its HEIF extension (denoted as ``PIL'' in our results), while VVC encoding was performed using VVenC~\cite{VVenC}. For Cheng's learned codec, we used the pretrained ``Anchor'' models provided by CompressAI~\cite{begaint2020compressai}. We further assessed generalization across two \acp{VLM}: LLaVA One Vision~\cite{llava_ov} (7B) and InternVL3~\cite{internvl3} (9B).

\begin{figure*}[!t]
    \centering
    \subfloat[]{%
        \includegraphics[width=0.29\textwidth]{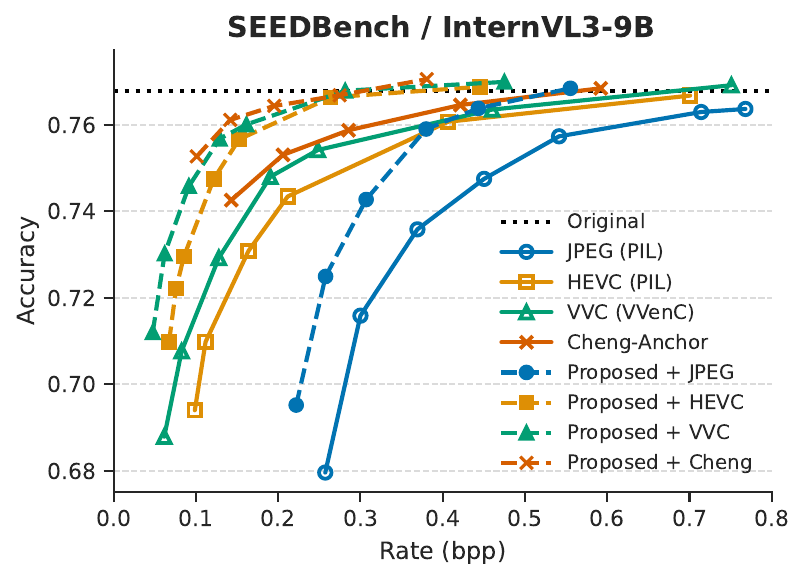}%
        \label{fig:seed_internvl}
    }
    \hfil
    \subfloat[]{%
        \includegraphics[width=0.29\textwidth]{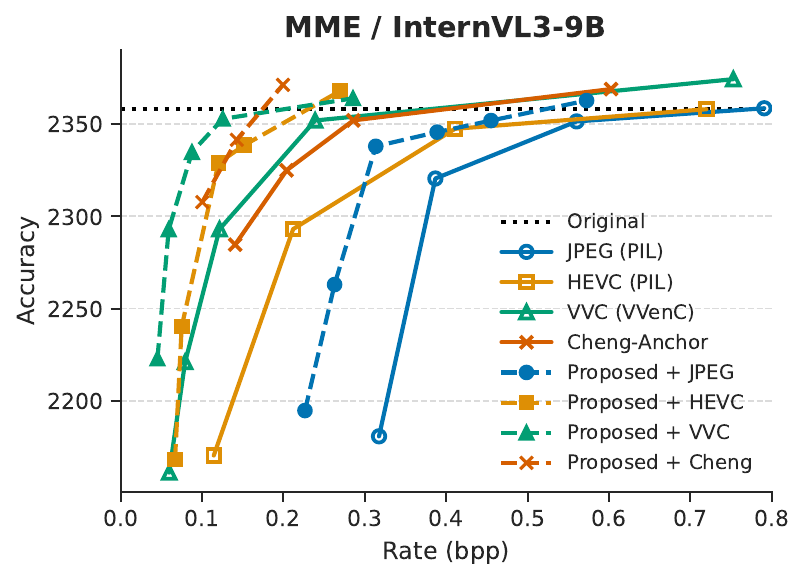}%
        \label{fig:mme_internvl}
    }
    \hfil
    \subfloat[]{%
        \includegraphics[width=0.29\textwidth]{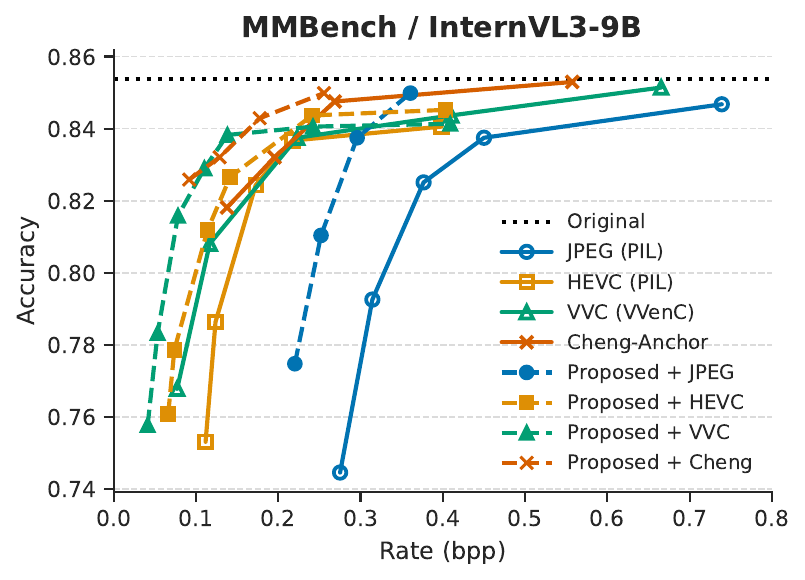}%
        \label{fig:mmbench_internvl}
    }


    \subfloat[]{%
        \includegraphics[width=0.29\textwidth]{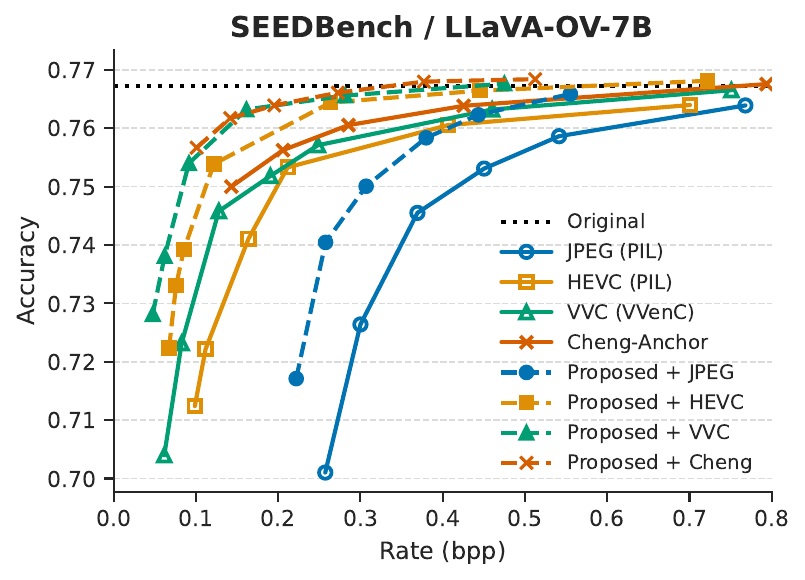}%
        \label{fig:seed_llava}
    }
    \hfil
    \subfloat[]{%
        \includegraphics[width=0.29\textwidth]{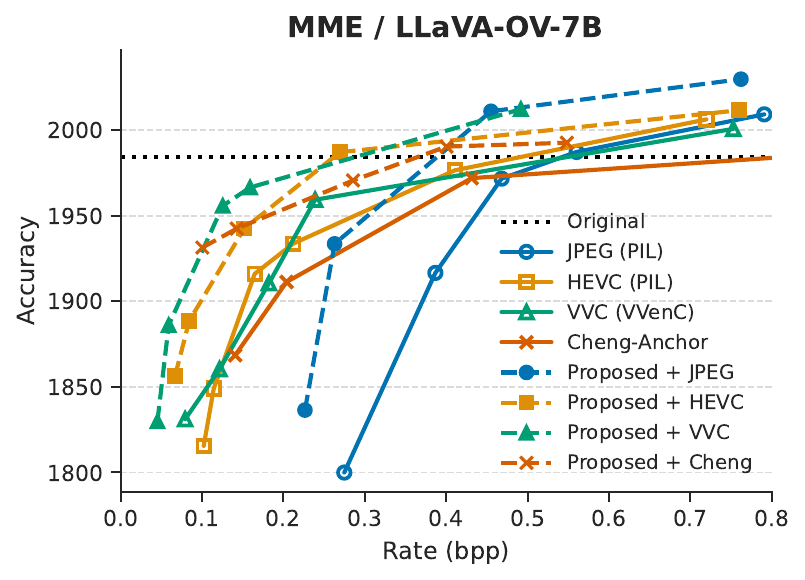}%
        \label{fig:mme_llava}
    }
    \hfil
    \subfloat[]{%
        \includegraphics[width=0.29\textwidth]{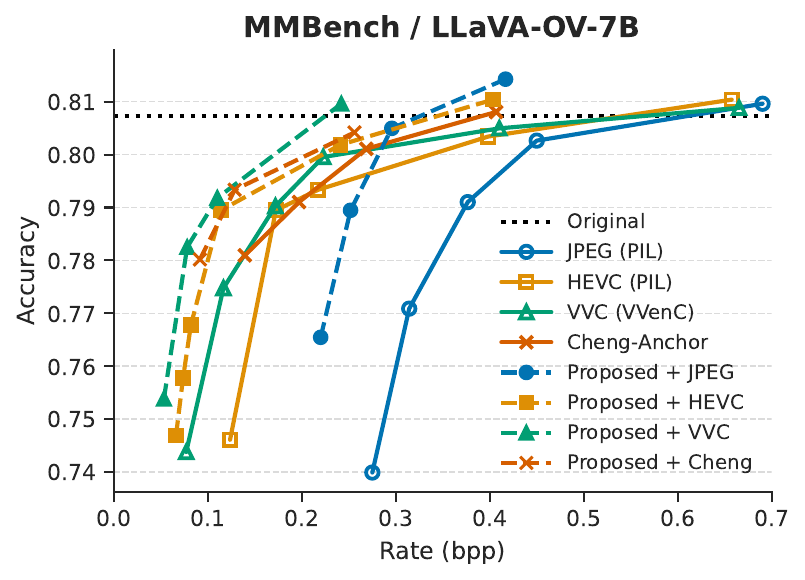}%
        \label{fig:mmbench_llava}
    }

    \caption{Rate–Accuracy curves for InternVL3-9B (first row) and LLaVA-OV-7B (second row) across SEEDBench, MME, and MMBench benchmarks.}
    \label{fig:rate_accuracy}
    \vspace{-0.1mm}
\end{figure*}

Fig.~\ref{fig:rate_accuracy} presents six rate–accuracy plots corresponding to the three datasets and two \acp{VLM}. Each plot includes rate–accuracy curves for the four pure compression methods (solid lines) alongside their prefiltering-enhanced counterparts (dashed lines). In all cases, integrating our prefiltering module consistently improves performance compared with using the encoders alone. The different points on each curve correspond to varying the quality factor or quantization parameter of the underlying codec.

Table~\ref{tab:bd_rate_results} summarizes the Bj{\o}ntegaard-Delta~\cite{bjontegaard2001calculation} (BD-rate) values between each prefiltered system and its corresponding standalone-compression baseline, using the latter as the anchor. The results show that applying our proposed prefiltering module reduces bitrate by 25.88\% to 48.78\% on average at the same level of \ac{VLM} accuracy. As expected, the smallest gain is observed with the simplest codec (JPEG), while the largest gain occurs with the most advanced codec (VVC), likely because VVC can more effectively exploit the low-pass characteristics introduced in the smoothed regions.

\begin{table}[!t]
\centering
\sisetup{detect-weight, mode=text} 
\caption{
    BD-Rate (\%) comparing our prefiltered approach against the corresponding baseline (same codec without prefiltering). Negative values indicate bitrate savings.
}
\label{tab:bd_rate_results}
\resizebox{\columnwidth}{!}{
\begin{tabular}{
    l c                  
    S[table-format=-2.1] 
    S[table-format=-2.1] 
    S[table-format=-2.1] 
    S[table-format=-2.1, 
      table-column-width=1.2cm] 
}
\toprule
\multirow{2.5}{*}{\textbf{VLM}} & \multirow{2.5}{*}{\shortstack{\textbf{Compression} \\ \textbf{Method}}} & \multicolumn{4}{c}{\textbf{BD-Rate} (\%) $\downarrow$} \\

\cmidrule(lr){3-6} 
 & & {MMBench} & {MME} & {SEEDBench} & {\textbf{Average}} \\
\midrule

\multirow{4}{*}{\textbf{InternVL}} 
 & JPEG & -29.30 & -24.55 & -23.78 & \textbf{-25.88} \\
 & HEVC & -30.60 & -51.93 & -48.54 & \textbf{-43.69} \\
 & VVC  & -38.27 & -49.02 & -52.05 & \textbf{-46.45} \\
 & Cheng~\cite{Cheng_2020_CVPR}  & -29.65 & -45.64 & -53.50 & \textbf{-42.93} \\
\cmidrule(lr){1-6} 

\multirow{4}{*}{\textbf{LLaVA}} 
 & JPEG & -33.22 & -33.83 & -25.66 & \textbf{-30.90} \\
 & HEVC & -37.99 & -38.87 & -46.74 & \textbf{-41.20} \\
 & VVC  & -42.13 & -53.47 & -50.74 & \textbf{-48.78} \\
 & Cheng~\cite{Cheng_2020_CVPR}  & -34.96 & -42.54 & -55.27 & \textbf{-44.26} \\
\bottomrule
\end{tabular}
}
\end{table}

\subsection{Ablation Study}
\label{subsec:ablation}
To evaluate the contribution of each component in our prefiltering module, we conducted an ablation study in which individual elements were removed or replaced with alternative configurations. The tested variants include:
(i) our full proposed method using the 57M-parameter TinyCLIP model, denoted ``Ours-57M'';
(ii) a version without the TinyCLIP component, denoted ``w/o TinyCLIP'', where all tiles are assigned a uniform score of 1 and every block is filtered with a fixed value of $\sigma_{\mathbf{1}} = 0.2$;
(iii) a version without overlapping tiles in the Tile Extraction module ($stride = t_{size}$ always), denoted ``NoOverlap'';
(iv) a version without the Prompt Preprocessing module, denoted ``w/o PP''; and
(v) two additional versions of our full pipeline using alternative TinyCLIP models with 23M\footnote{\texttt{TinyCLIP-ViT-8M-16-Text-3M}} and 120M\footnote{\texttt{TinyCLIP-auto-ViT-63M-32-Text-31M}} parameters, denoted ``Ours-23M'' and ``Ours-120M'', respectively.
Appendix B provides alternative whole-image filtering baselines for comparison.

\begin{figure}[t]
\centerline{\includegraphics[width=0.85\columnwidth]{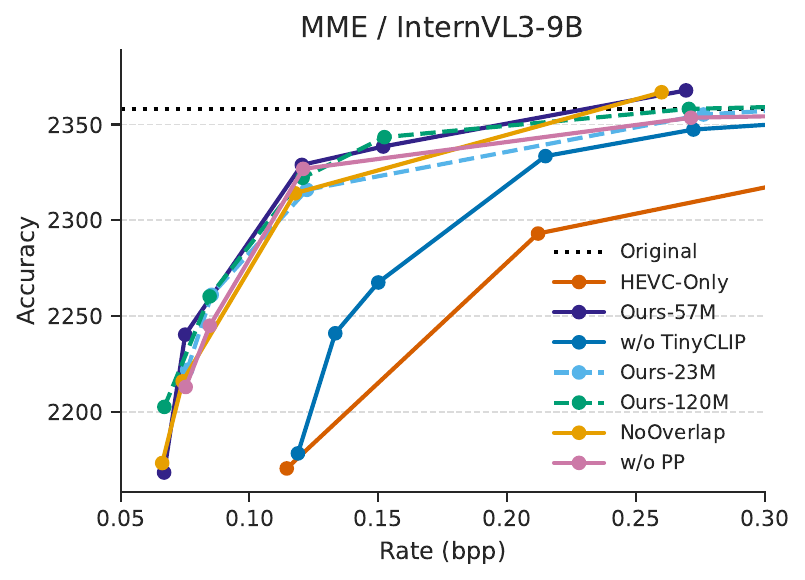}}
\caption{Rate–accuracy curves for InternVL3-9B on the MME dataset, illustrating the impact of different components of our method.}
\label{fig:ablation}
\end{figure}

All experiments here use the HEVC codec. A sample rate–accuracy plot for these variants, evaluated on the MME dataset with the InternVL3 model is shown in Fig.~\ref{fig:ablation}. The variant labeled ``HEVC-Only'' corresponds to the baseline without any prefiltering module. The corresponding average BD-rate values across all three datasets are reported in Table~\ref{tab:ablation_study} for both InternVL and LLaVA models, using our full proposed method with 57M parameters as the anchor.

Based on these results, and consistent with the findings in the previous section, removing the entire prefiltering pipeline leads to an approximately 77\% increase in bitrate. The most critical component of the system is the TinyCLIP model, which serves as the core mechanism for identifying prompt-relevant and irrelevant image regions. Eliminating TinyCLIP results in roughly a 70\% bitrate increase. Interestingly, this is still slightly better than the baseline (no prefiltering), likely because the target VLMs are somewhat robust to mild, uniform Gaussian smoothing. Although the gains from overlapping tiles (which refine the score map granularity) and prompt preprocessing are smaller, both components still contribute meaningfully to overall compression efficiency. Additionally, reducing the TinyCLIP size leads to about a 6\% increase in bitrate, and increasing its size slightly worsens performance too. This trend could be a sign of reaching a saturation point in TinyCLIP’s contribution, implying that the 57M model is already optimal for our setting.

\begin{table}[!t]
\centering
\sisetup{
    detect-weight, 
    mode=text, 
    retain-explicit-plus=true
} 
\caption{
    Average BD-Rate (\%) across SeedBench, MME, and MMBench for method ablations, using ``Ours-57M'' as the anchor.
}
\label{tab:ablation_study}
\begin{tabular}{
    l                     
    S[table-format=+2.2]  
    S[table-format=+2.2]  
    S[table-format=+2.2]  
}
\toprule
Method Variant & {InternVL} & {LLaVA} & {\textbf{Average}} \\
\midrule
\textbf{Ours-57M} & \textbf{0} & \textbf{0} & \textbf{0} \\
\midrule
HEVC-Only       & +82.40 & +72.73 & +77.57 \\
\midrule
w/o TinyCLIP    & +71.41 & +70.42 & +70.91 \\
\midrule
Ours-23M        & +7.13 & +4.51 & +5.82 \\
\midrule
Ours-120M       & +6.69 & -2.06 & +2.31 \\
\midrule
NoOverlap       & +4.12 & +6.15 & +5.13 \\
\midrule
w/o PP          & +7.65 & +0.04 & +3.85 \\
\bottomrule
\end{tabular}
\end{table}

\section{Conclusion}
\label{sec:conc}
In this work, we addressed the inefficiency of using human-centric codecs for transmitting visual data to cloud-based \acp{VLM}. We introduced a novel, prompt-guided prefiltering module that leverages a lightweight TinyCLIP model to dynamically identify and preserve task-relevant image regions while smoothing irrelevant areas. Unlike existing \ac{ICM} approaches that assume a fixed downstream task, our method adapts to the open-ended nature of VLM prompts, ensuring that critical visual details are retained for the specific query at hand. Being codec-agnostic, our module can be seamlessly integrated as a plug-and-play preprocessor before any image encoder, without requiring modifications to the edge hardware or the cloud-based decoder. Our experiments on MMBench, SEEDBench, and MME benchmarks demonstrate that our approach achieves significant bitrate savings of approximately 28\% for JPEG, 42\% for HEVC, 47\% for VVC standard codecs, and 44\% for a learned image codec by Cheng \textit{et al.} while maintaining \ac{VQA} accuracy. These results underscore the potential of prompt-aware prefiltering as a practical and efficient solution for the growing field of edge-to-cloud multimodal intelligence.

\bibliographystyle{IEEEtran}
\bibliography{refs}

\appendices

\section{Latency and Complexity}

As discussed in Section III, deploying the prefiltering module is highly practical for modern edge devices. The TinyCLIP model requires only 23M or 57M parameters, making it more lightweight than state-of-the-art learned image codecs. To empirically demonstrate that this module is justified in terms of computational complexity and end-to-end latency, we measured GPU memory usage and inference time across 1,000 images from the SEEDBench dataset. All evaluations were conducted on a consumer-grade laptop equipped with an NVIDIA GeForce RTX 3050 Ti GPU (4 GB VRAM).

At peak capacity, the full inference pipeline with TinyCLIP-57M model consumes a modest 305 MB of GPU memory. Furthermore, the average processing latency of the prefiltering module is $35.80 \pm 0.45$~ms (95\% confidence interval). Crucially, this minor processing overhead is more than offset by the subsequent reduction in transmission latency for typical use cases. For example, transmitting a 50-megapixel image (typical for modern smartphones) compressed at 0.1 bpp over a standard 10 Mbps upload link requires approximately 500 ms. Because our prefiltering module achieves a $\sim$42\% reduction in bitrate, it saves roughly 210~ms in transmission time, easily compensating for the initial $\sim$36~ms prefiltering delay.

\section{Alternative Baselines}

\begin{figure}[tb] \centering
    \includegraphics[width=\columnwidth]{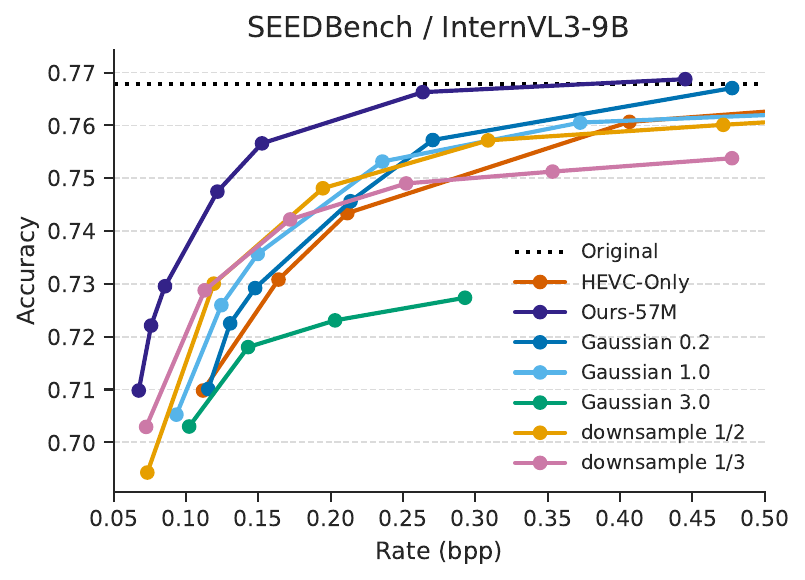}
    \caption{Rate–accuracy curves for InternVL3-9B on SEEDBench dataset} \label{fig:supp-abb}
\end{figure}

In this section, we compare our proposed method against several simpler baselines using the same experimental settings as Section IV-B (Ablation Study), with HEVC as the underlying codec. These baselines rely on whole-image processing strategies, specifically global Gaussian filtering with $\sigma \in \{0.2, 1.0, 3.0\}$ and image downsampling using bilinear interpolation with ratios $\in \{1/2, 1/3\}$. Fig.~\ref{fig:supp-abb} illustrates the rate-accuracy curves for these baselines, alongside our 57M-parameter proposed method (``Ours-57M'') and the HEVC-Only baseline. Additionally, Table~\ref{tab:supp-baselines} presents the corresponding average BD-Rate values across the three datasets. It is worth noting that the ``w/o TinyCLIP'' configuration from our Ablation Study is effectively global Gaussian filtering with $\sigma=0.2$; without the TinyCLIP module, all tiles receive a uniform score of 1, causing all blocks to be filtered with a constant $\sigma_{\mathbf{1}} = 0.2$.

As the results show, while these global baselines are straightforward to implement and successfully improve compression efficiency compared to standard HEVC, they fall significantly short of our proposed method. Introducing low-intensity global Gaussian filtering provides an initial gain over the HEVC-Only baseline. However, aggressively increasing the smoothing factor (e.g., $\sigma=3.0$) does not yield further compression benefits and instead leads to a severe drop in task accuracy. Similarly, while increasing $\sigma$ from 0.2 to 1.0 yields overall bitrate savings, it begins to degrade performance at higher bitrates. This pattern also holds true for downsampling: while an initial downsampling ratio provides bit savings over pure HEVC, increasing the ratio further results in noticeable performance penalties at high bitrates.

\begin{table}[!t]
\centering
\sisetup{
    detect-weight, 
    mode=text, 
    retain-explicit-plus=true
}
\caption{
    Average BD-Rate (\%) across SeedBench, MME, and MMBench for different baselines, using ``Ours-57M'' as the anchor
}
\label{tab:supp-baselines}
\begin{tabular}{
    l                     
    S[table-format=+3.2]  
    S[table-format=+3.2]  
    S[table-format=+3.2]  
}
\toprule
Method Variant & {InternVL} & {LLaVA} & {\textbf{Average}} \\
\midrule
\textbf{Ours-57M} & \textbf{0} & \textbf{0} & \textbf{0} \\
\midrule
HEVC-Only       & +82.40 & +72.73 & +77.57 \\
\midrule
w/o TinyCLIP (Gaussian 0.2)    & +71.41 & +70.42 & +70.91 \\
\midrule
Global Gaussian ($\sigma=1.0)$ & +54.99 & +47.81 & +51.40 \\
\midrule
 Global Gaussian ($\sigma=3.0)$ & +155.89 & +115.65 & +135.77 \\
\midrule
Downsampling (1/2) & +49.50 & +61.26 & +55.38 \\
\midrule
Downsampling (1/3) & +39.44 & +51.15 & +45.29 \\
\bottomrule
\end{tabular}
\end{table}

\end{document}